\documentclass[11pt,a4paper]{paper}

\usepackage[a4paper,margin=1in]{geometry}
\usepackage{graphicx}
\usepackage{cite}
\usepackage{epsfig}
\usepackage{amssymb} 
\usepackage{lineno}
\usepackage{hyperref}

\newcommand\pubnumber{DPF2015-356}
\newcommand\pubdate{October 31, 2015}

\def\Title#1{\begin{center} {\Large #1 } \end{center}}
\def\Author#1{\begin{center}{ \sc #1} \end{center}}
\def\Address#1{\begin{center}{ \it #1} \end{center}}

\newcommand\pubblock{\rightline{\begin{tabular}{l} \pubnumber\\
         \pubdate  \end{tabular}}}
\newenvironment{Abstract}{\begin{quotation}  }{\end{quotation}}
\newenvironment{Presented}{\begin{quotation} \begin{center} 
             PRESENTED AT\end{center}\bigskip 
      \begin{center}\begin{large}}{\end{large}\end{center} \end{quotation}}

\def\beq{\begin{equation}}
\def\eeq#1{\label{#1}\end{equation}}
\def\eeqn{\end{equation}}

\def\beqa{\begin{eqnarray}}
\def\eeqa#1{\label{#1}\end{eqnarray}}
\def\eeqan{\end{eqnarray}}

\let\bar=\overbar

\def\Dslash{\not{\hbox{\kern-4pt $D$}}}
\def\dslash{\not{\hbox{\kern-2pt $\del$}}}

\def\msb{{\bar{\ssstyle M \kern -1pt S}}}

\def\support{\footnote{Now at MTA-ELTE Lend\"ulet CMS Particle and Nuclear Physics Group, E\"otv\"os Lor\'and University, Budapest, Hungary.}}

\begin{document}
\begin{titlepage}
\pubblock

\vfill
\Title{The Upgrade of the ATLAS Electron and Photon Triggers \\ towards LHC Run 2 and their Performance}
\vfill
\Author{ Gabriella P\'asztor\support \\ on behalf of the ATLAS Collaboration}
\Address{Carleton University, Ottawa, Canada}
\vfill
\begin{Abstract}
\noindent Electron and photon triggers covering transverse energies from 5 GeV to several TeV are essential for signal selection in a wide variety of ATLAS physics analyses to study Standard Model processes and to search for new phenomena. Final states including leptons and photons had, for example, an important role in the discovery and measurement of the Higgs particle. Dedicated triggers are also used to collect data for calibration, efficiency and fake rate measurements. The ATLAS trigger system is divided in a hardware-based (Level 1) and a software based High-Level Trigger (HLT), both of which were upgraded during the long shutdown of the LHC in preparation for data taking in 2015. The increasing luminosity and more challenging pile-up conditions as well as the higher center-of-mass energy demanded the optimisation of the trigger selections at each level, to control the rates and keep efficiencies high. To improve the performance, multivariate analysis techniques were introduced at the HLT. The evolution of the ATLAS electron and photon triggers and their performance is presented, including initial results from the early days of the LHC Run 2 operation.
\end{Abstract}
\vfill
\begin{Presented}
DPF 2015\\
The Meeting of the American Physical Society\\
Division of Particles and Fields\\
Ann Arbor, Michigan, August 4--8, 2015\\
\end{Presented}
\vfill
\end{titlepage}
\def\thefootnote{\fnsymbol{footnote}}
\setcounter{footnote}{0}

\section{Introduction}

Electron and photon triggers play an essential role at the LHC. They select, for example, events containing $W \rightarrow e\nu$ and $Z \rightarrow ee$ decays, processes that are important on their own right to test the Standard Model and to calibrate the experimental apparatus but can also be part of the decay of heavier objects and thus help us in our quest to find new phenomena. Indeed, these triggers enabled the ATLAS collaboration in 2012 to discover the Higgs boson via its decays to Z, W and photon pairs ($H \rightarrow ZZ^* \rightarrow 4\ell$, $H \rightarrow WW^* \rightarrow \ell\nu\ell\nu$, and 
$H \rightarrow \gamma\gamma$) and might also lead us to other new particles, such as new gauge bosons ($Z^\prime \rightarrow ee$) or excited graviton states ($G_\mathrm{KK} \rightarrow \gamma\gamma$).

The increased energy and luminosity of the LHC in Run2 necessitated the upgrade of the trigger system to keep event rates under control while maintaining high efficiencies for interesting processes. The ATLAS collaboration developed an ambitious upgrade program and its first stage was successfully completed during the long shutdown of the LHC during 2013 $-$ 2015. In the following sections the upgraded electron and photon trigger system and its performance in the first 2015 proton -- proton collision data is presented. 

\section{Electron and photon triggers in ATLAS}

The ATLAS detector is described in Ref.~\cite{ATLAS}. Electron and photon reconstruction~\cite{ATLASElectron,ATLASPhoton} relies primarily on the finely segmented calorimeter system and on the inner tracking detectors based on Silicon pixel and strip detectors in the inner-most part, followed by a Transition Radiation Tracker (TRT) providing also electron -- hadron separation via the detection of transition radiation photons.

The trigger system~\cite{ATLASTrigger} reduces the event rate to be recorded to about 1 kHz from the LHC beam crossing rate of 40 MHz. It is based on the Region-of-Interest concept in which the software-based high-level trigger (HLT) reconstruction is seeded by the level-1 (L1) objects provided by the hardware trigger. In particular, electron and photon trigger~\cite{ATLASEgammaTrigger} decisions always start from the input of the level-1 calorimeter trigger that is based on trigger towers of $0.1 \times 0.1$ size in the pseudorapidity ($\eta$) -- azimuthal angle ($\phi$) plane. 

\begin{figure}[h]
\includegraphics[width=0.7\textwidth, trim= 50mm 100mm 30mm 100mm, clip]{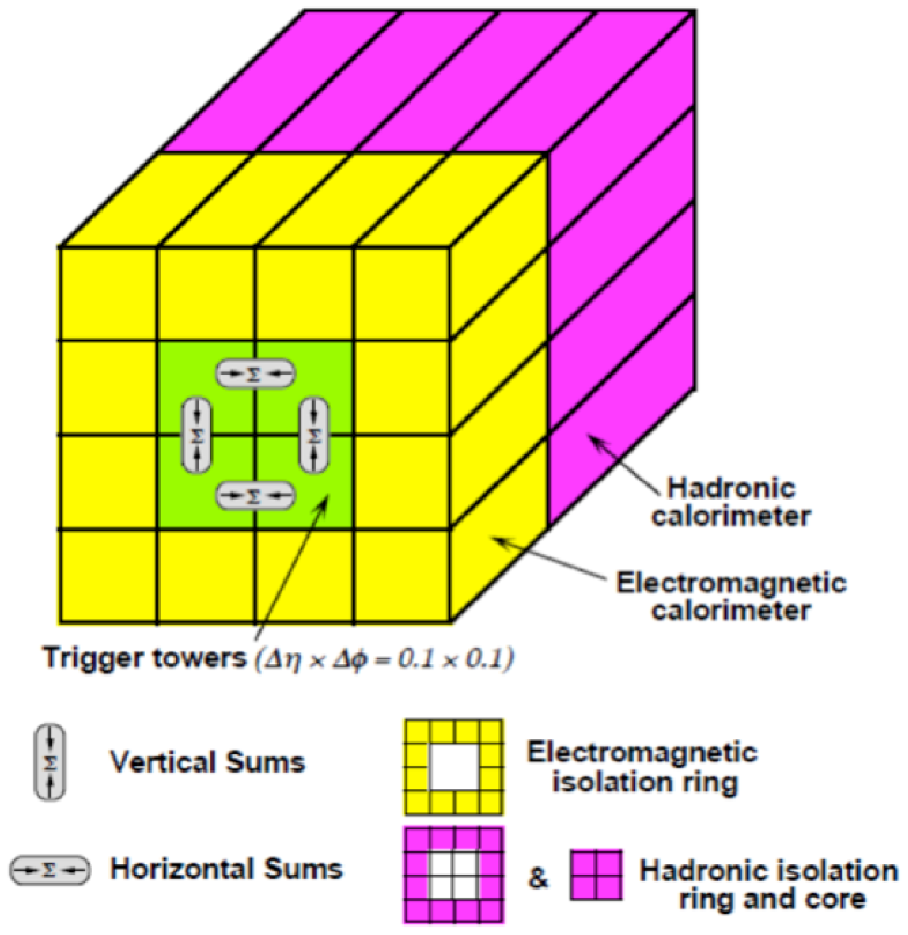}
\hspace*{-0.5cm}
\includegraphics[width=0.7\textwidth, trim= 27mm 117mm 25mm 115mm, clip, angle=90]{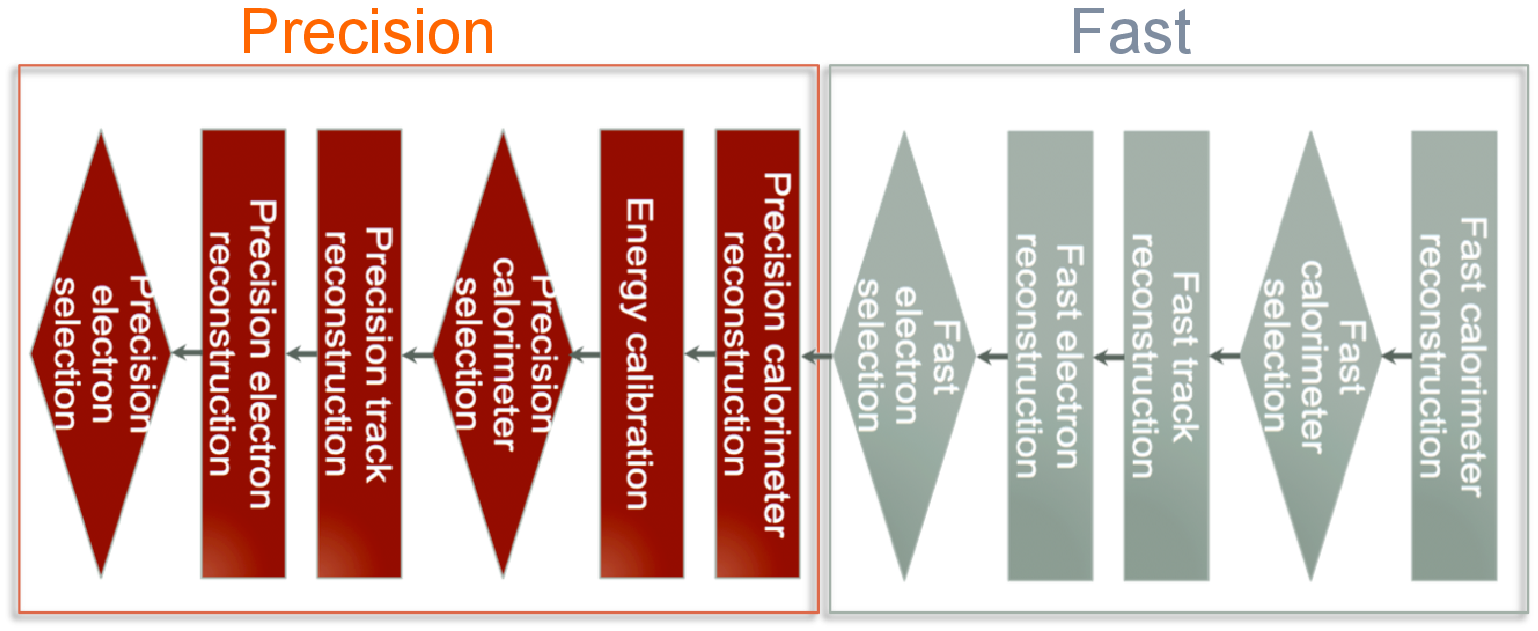}
\caption{(left) The L1 calorimeter cluster for electron and photon triggers. (right) The HLT trigger algorithm sequence for electron triggers.}
\label{fig:ATLASTrigger}
\end{figure}

The electromagnetic cluster reconstruction at L1 uses a sliding-window algorithm to find local energy maxima and provides the cluster energy collected in 2x2 trigger towers in the electromagnetic (EM) calorimeter. To discriminate against hadron jets, it also computes the energy sum in the isolation ring formed by the surrounding 12 towers in the EM calorimeter as well as the hadronic core energy behind the 2x2 EM cluster, as illustrated on the left of Figure~\ref{fig:ATLASTrigger}.

Already in the LHC Run-1 in 2010 $-$ 2013, the L1 EM cluster transverse energy ($E_\mathrm{T}$) threshold was pseudorapidity dependent to take into account the energy loss in the detector material before the calorimeter. The threshold could be set by $\Delta \mathrm{E_T} \sim 1$~GeV precision and with $\Delta\eta = 0.4$ granularity. For the main unprescaled EM triggers a veto on hadronic core energy above 1 GeV was also typically required.

The upgrade of the L1 calorimeter trigger during the long LHC shut-down in 2013 $-$ 2015 brought many improvements. The new Multi Chip Module (nMCM) in the Pre-Processor responsible for the signal processing, now features a noise autocorrelation filter to achieve better energy resolution as well as dynamic pedestal correction. The firmware upgrade of the Cluster Processor Module (CPM) allows the definition of five $E_\mathrm{T}$-dependent electromagnetic and/or hadronic core isolation selections with a precision of $\Delta \mathrm{E_T} \sim 0.5$~GeV. Moreover the new Extended Common Merger Module (CMX) doubles the number of $E_\mathrm{T}$ thresholds to 16. The threshold values can now be set by 
$\Delta\eta = 0.1$ granularity bringing a better trigger efficiency uniformity in pseudorapidity.

Tracking information is first used at the HLT which defines \textsl{photons} as electromagnetic energy clusters with no requirement on a matching track and \textsl{electrons} as energy clusters matched to reconstructed charged particle tracks with a transverse momentum above 1 GeV and having a minimum number of hits in the inner Silicon tracking devices. 

Several changes were introduced at the HLT. The algorithm sequence is shown on the right of Figure~\ref{fig:ATLASTrigger} for electron triggers. As calorimeter reconstruction is less resource intensive it precedes the tracking step. Photon triggers operate in a similar fashion but are simpler as only calorimeter reconstruction and selection is applied. The previously two-level HLT reconstruction is merged to run on a single computer farm and have now a common data preparation for the fast and precision online reconstruction steps. The initial fast reconstruction helps to reduce the event rate early. In Run 2, the fast calorimeter reconstruction and selection can be skipped, but fast track reconstruction is always run for electron triggers and seeds precision tracking. The final online precision reconstruction is improved and uses offline-like algorithms as much as possible. In particular a new electron and photon energy calibration and a new electron identification are introduced online, both based on multivariate analysis techniques.

\section{Trigger performance}

\subsection{Energy resolution}

\begin{figure}
\centering
\includegraphics[width=0.49\textwidth,height=0.355\textwidth, trim= 0mm 0mm 0mm 0mm, clip]{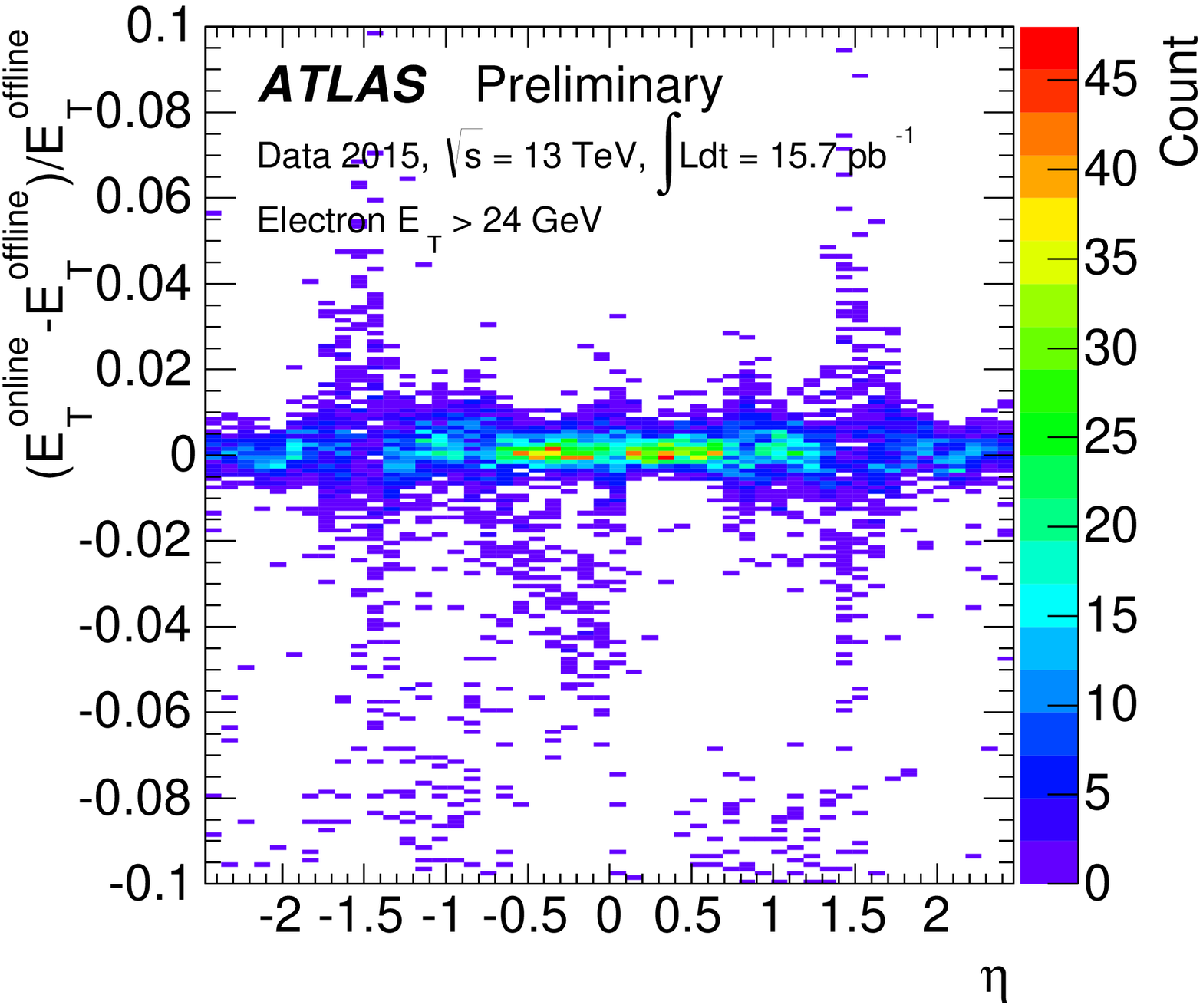}
\includegraphics[width=0.49\textwidth]{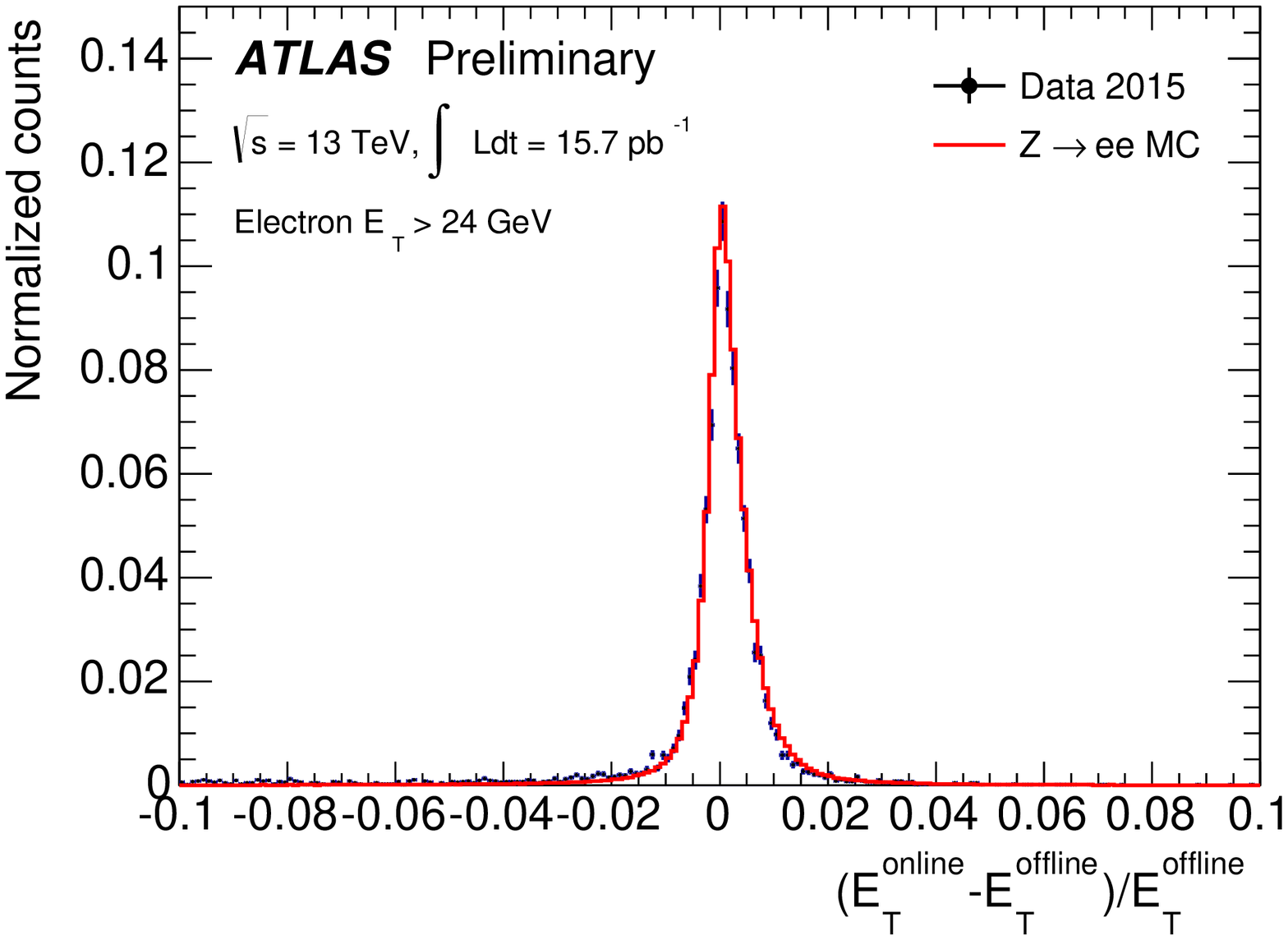}
\caption{Electron energy resolution online with respect to the offline reconstruction~\cite{ATLASEgammaTriggerPrelim}.}
\label{fig:Calibration}
\end{figure}

Cluster energy calibration corrects the measured energy for losses upstream of the calorimeter as well as for lateral and longitudinal energy leakage outside the calorimeter cluster. The online reconstruction uses a simplified version of the offline method relying on boosted decision trees to determine the correction factors. Separate calibration is used for electrons and photons, however photons are not separated to converted and unconverted categories at the HLT which is a major source of the remaining differences with respect to offline reconstruction. Figure~\ref{fig:Calibration} shows the energy resolution for electrons with respect to the offline calibration as a function of pseudorapidity (on the left) and compares the measured resolution to the expectation from Monte Carlo simulation (on the right). While the resolution is excellent in most of the pseudorapidity range, it worsens considerably in the transition region between the barrel and endcap electromagnetic calorimeters at $|\eta|=1.37-1.52$ where a large amount of material is present upstream of the calorimeter. 

\subsection{Rate and efficiency} 
 
Photon identification in ATLAS relies on shower-shape information from the calorimeter system and is based on rectangular cuts optimised in different pseudorapidity regions. While offline converted and unconverted photon candidates are separated and have different identification selections, online no attempt is made for conversion reconstruction and the looser selection cuts are applied from the two optimisations.

For electron identification to improve the purity of the triggered data sample, a new likelihood-based approach was adopted online which was successfully used offline already in Run 1. It uses input from calorimeter shower-shapes, tracking, track -- cluster matching and a new electron probability derived from transition radiation information measured in the TRT. Based on measurements in 2012 data using offline reconstruction, the likelihood-based selection provides about a factor two improvement in background rejection for the same signal efficiency with respect to the optimised cut-based electron selection. The largest difference between the online and the offline implementation originates from the lack of dedicated bremsstrahlung correction with the Gaussian Sum Filter method online.
 
\begin{figure}
\centering
\includegraphics[width=0.49\textwidth]{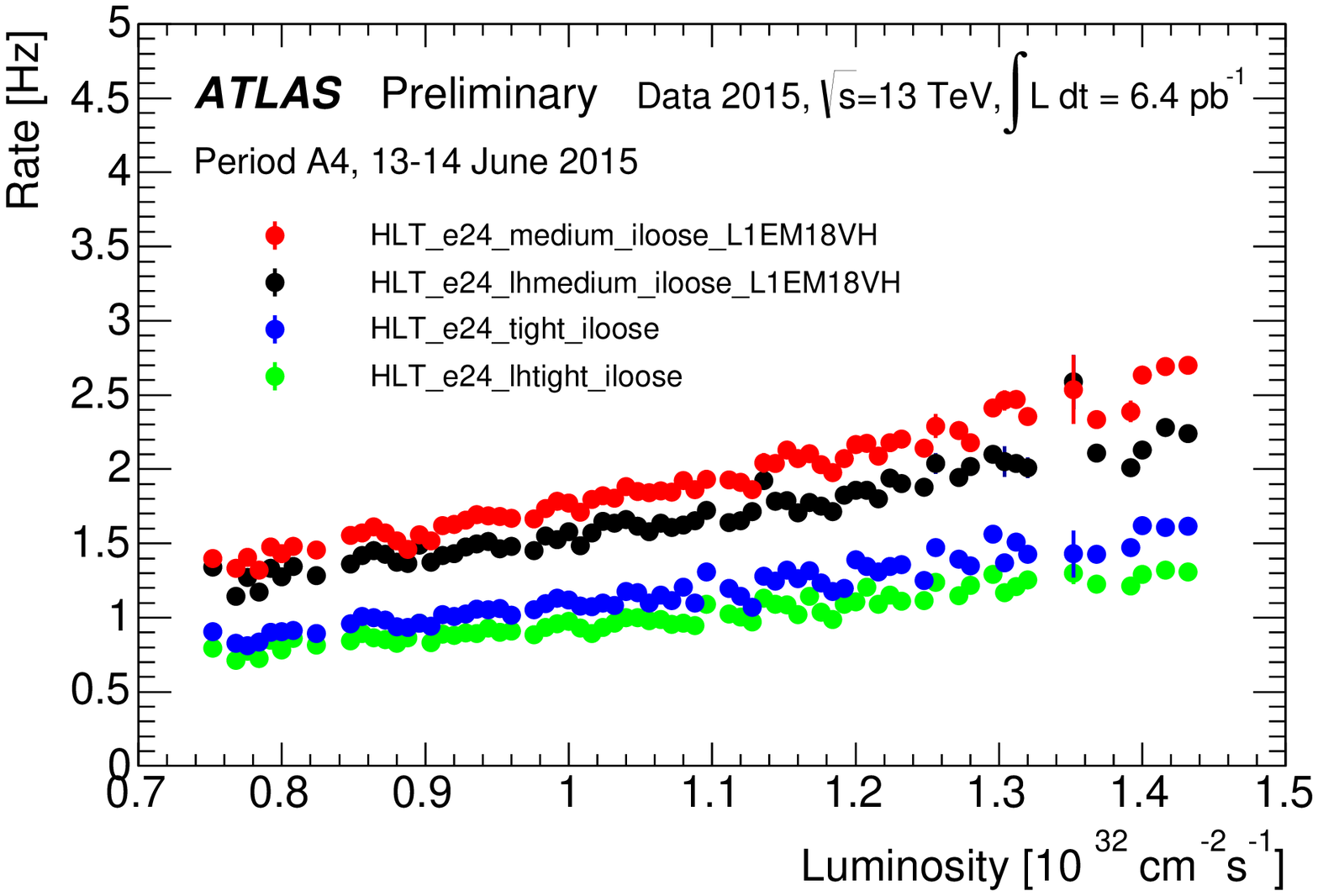}
\includegraphics[width=0.49\textwidth]{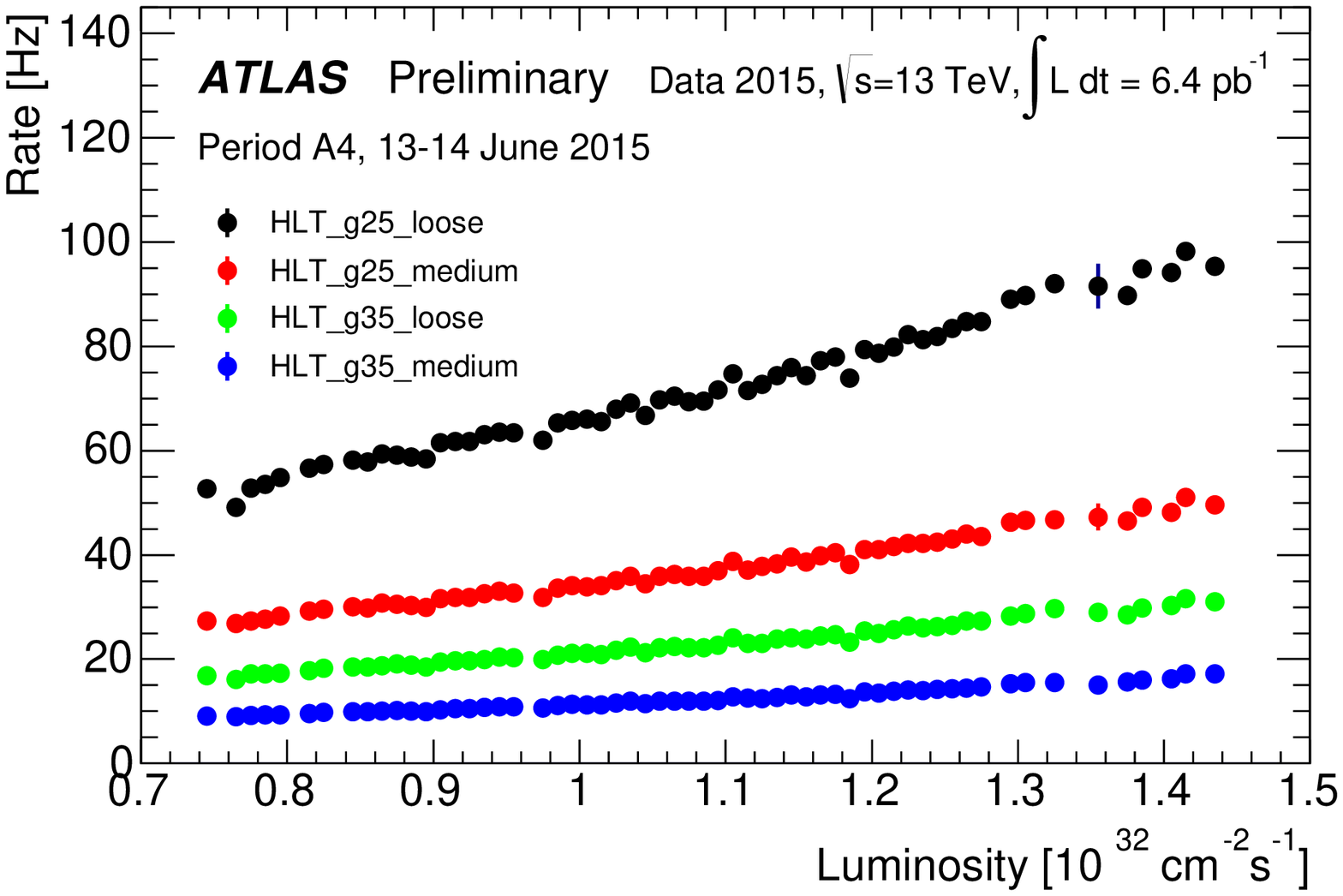}
\caption{Electron and photon trigger rates in early 2015 data taking~\cite{ATLASEgammaTriggerPrelim}.}
\label{fig:Rates}
\end{figure}

The ATLAS HLT strategy in Run 2 aims to keep online transverse energy thresholds at the Run 1 level (e.g. 24 GeV of transverse energy for single electron triggers) as long as possible by tightening the L1 and HLT selections gradually as the instantaneous luminosity increases. The trigger rates of different single electron and photon triggers\footnote{
ATLAS trigger names follow a well-defined convention. Photon / electron triggers start with a "g" / "e" followed by the transverse energy threshold in GeV. The identification selection is also given (e.g. \textit{lhloose, loose, lhmedium, medium, lhtight, tight}) as well as the presence of isolation cut at the HLT, if any (e.g. \textit{iloose} for electron triggers means that within an isolation  cone of R=0.2 the track momentum sum can not be more than 10\% of the electron transverse energy). The "L1" seed is also given if it is not the default for a given HLT threshold (e.g. in e24\_lhmedium\_iloose\_L1EM18VH, "EM18" indicates that an EM cluster with at least 18 GeV is required at L1, "V" indicates that the threshold is modified as a function of pseudorapidity to correct the effect of material before the calorimeter, and "H" ("I") that a hadronic (electromagnetic) isolation selection is requested). If multiple objects are requested the multiplicity is also given (e.g. 2e17\_lhloose). If several different objects are required they listed after each other (e.g. g35\_loose\_g25\_loose).} 
are shown on Figure~\ref{fig:Rates}. By tightening the photon selection from \textit{loose} to \textit{medium}, almost a factor two rate reduction is achieved for negligible loss of efficiency. Similarly, a rate reduction of about 45\% is observed when moving the likelihood-based electron selection from \textit{lhmedium} to \textit{lhtight}, also adding an EM isolation criteria at L1. The likelihood selections have about 20\% lower rates than the cut-based ones of similar tightness. For example, the \textit{lhmedium} selection is not only tuned to be about 6\% more efficient for true reconstructed electrons than its cut-based \textit{medium} counterpart but also results in a 20\% lower rate.

\begin{figure}
\centering
\includegraphics[width=0.49\textwidth]{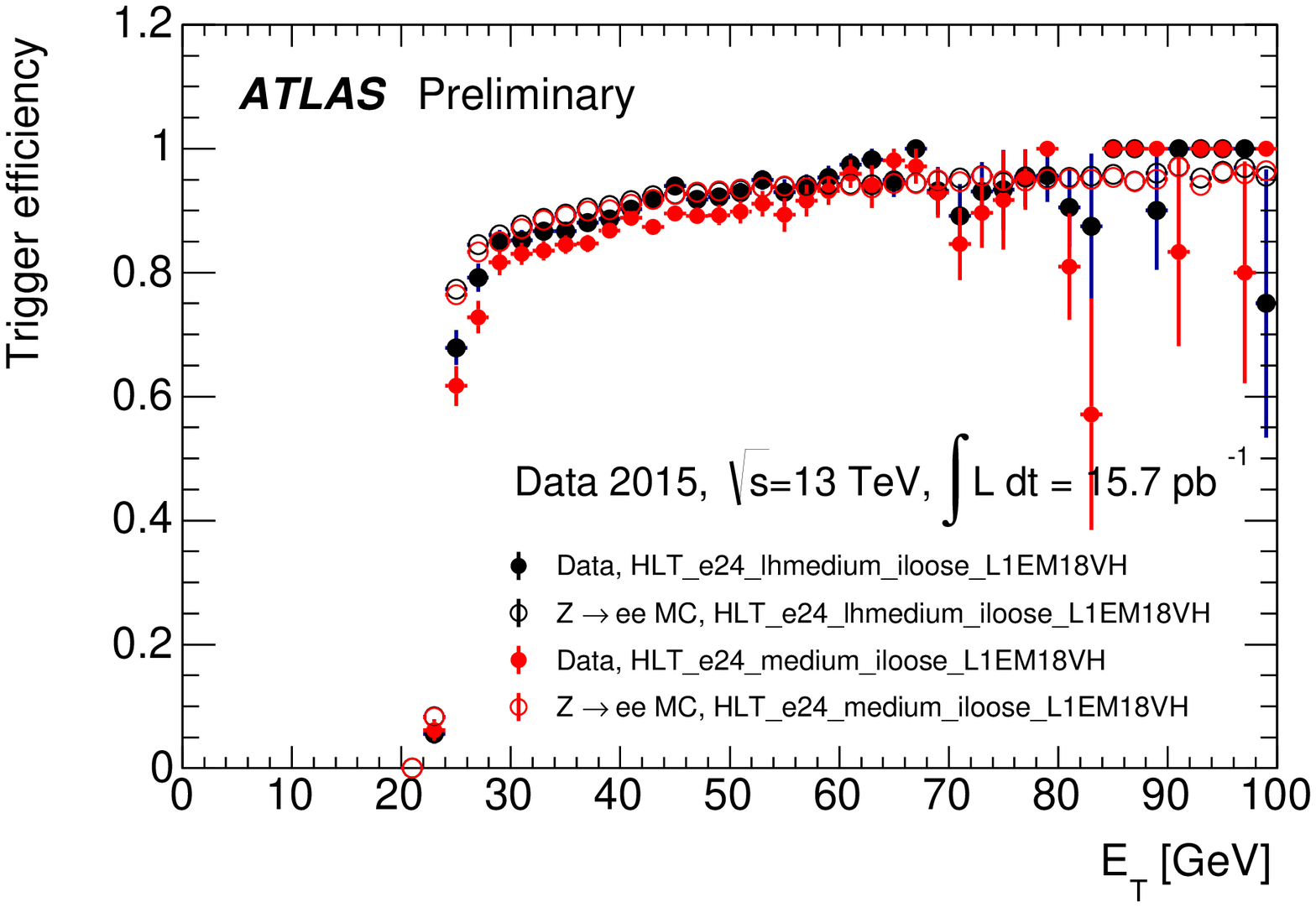}
\includegraphics[width=0.49\textwidth]{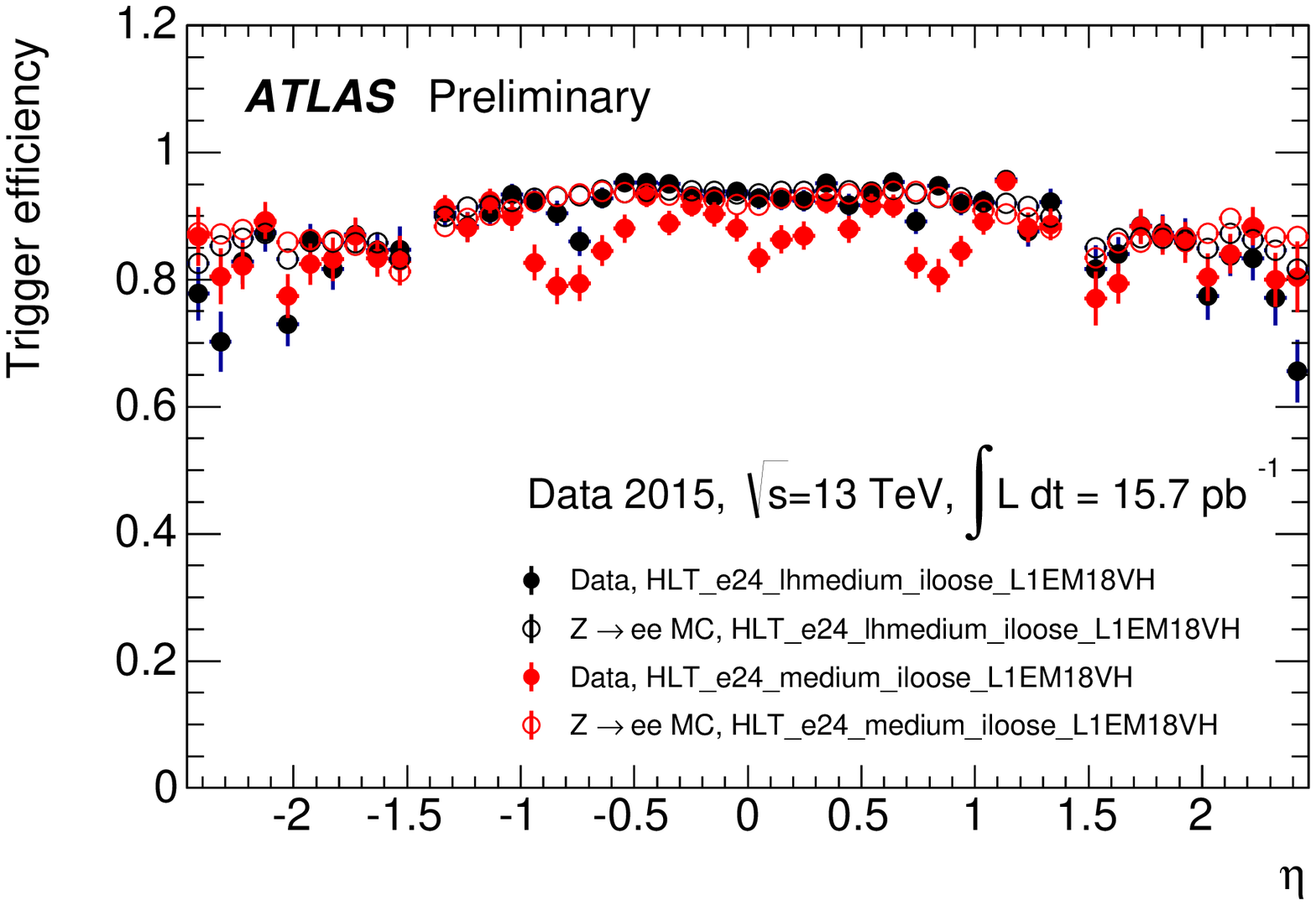}
\caption{Electron trigger efficiency with respect to the offline selection of the same identification level in early 2015 data taking~\cite{ATLASEgammaTriggerPrelim}. The efficiencies were measured with a tag-and-probe method using $Z \rightarrow ee$ decays with no background subtraction applied.}
\label{fig:EleEffi}
\end{figure}

Moreover, not only the performance but also the agreement between data and MC simulation is superior for likelihood triggers. This is visible in Figure~\ref{fig:EleEffi} which shows the single electron trigger efficiency with respect to offline reconstructed electrons passing the same identification level for cut-based and likelihood triggers comparing measurements in data and simulation as a function of electron transverse energy (on the left) and pseudorapidity (on the right). 

Detailed studies primarily on Run 1 data revealed the main sources of efficiency loss. The L1 energy resolution contributes significantly close to the transverse energy threshold. Both fast and precision HLT algorithms introduce inefficiencies predominantly due to tracking related selections. At high transverse energies, track isolation losses become significant which is recovered by introducing a non-isolated electron trigger with 60 GeV threshold. The single electron trigger efficiency in 2012 thus reached 95\% in most of the transverse energy -- pseudorapidity plane~\cite{ATLASEgammaTriggerPrelim}. It was measured with 0.1\% precision in the barrel region of $|\eta|<1.37$ for electrons with 30 $-$ 50 GeV transverse energy and up to 1\% elsewhere, using a tag-and-probe technique selecting $Z \rightarrow ee$ decays. 
 
\begin{figure}
\centering
\includegraphics[width=0.49\textwidth]{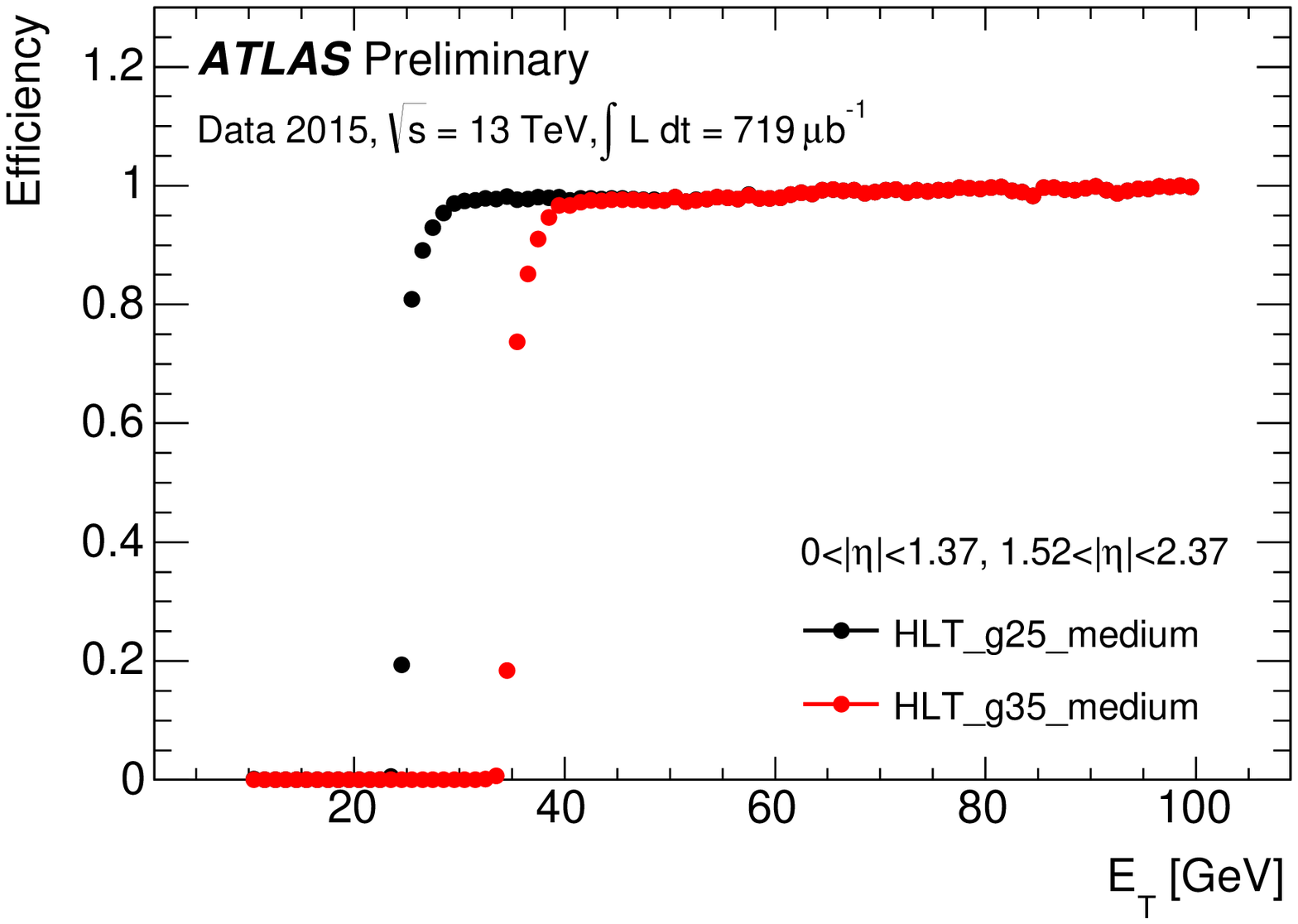}
\includegraphics[width=0.49\textwidth]{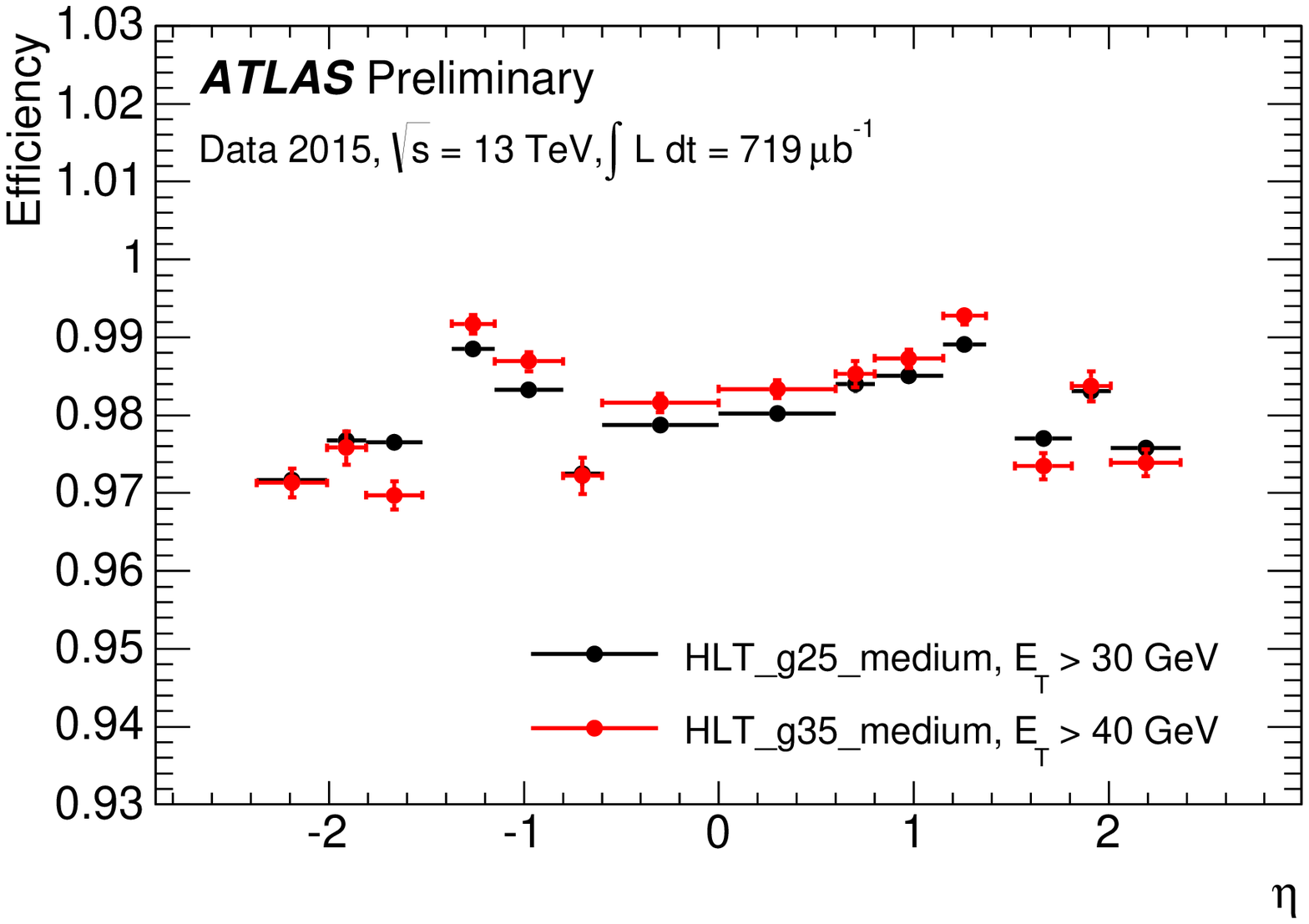}
\caption{Photon trigger efficiency with respect to offline tight selection in early 2015 data taking~\cite{ATLASEgammaTriggerPrelim}. The efficiency is measured using events recorded with a level-1 trigger requiring an electromagnetic cluster with 7 GeV transverse energy.
No background subtraction applied.}
\label{fig:PhoEffi}
\end{figure}

The efficiency of single photon triggers with respect to tight offline selection is illustrated on Figure~\ref{fig:PhoEffi} as a function of photon transverse energy (on the left) and pseudorapidity (on the right). The efficiency plateau is reached about 5 GeV above the transverse energy threshold.
As no background subtraction is applied in these early measurements, some of the inefficiencies are due to the impurity of the sample. As a comparison, in 2012 data the main di-photon trigger efficiency (HLT\_g35\_loose\_g25\_loose) was measured to be 99.50$\pm$0.15\%~\cite{ATLASEgammaTriggerPrelim}. 

\section{Outlook}

Many improvements were made to the ATLAS trigger system and to the online electron and photon reconstruction and identification in preparation for LHC Run 2 to keep the trigger thresholds at (or as close as possible to) the Run 1 levels in spite of the expected L1 rate increase of about a factor 5 due to the higher center-of-mass energy and the foreseen increase in instantaneous luminosity. 
During the 2015 data taking the single and di-electron trigger transverse energy thresholds could thus been kept at 24 GeV and 12 GeV, respectively, while single and di-photon triggers operated with 120 GeV and asymmetric (25 GeV, 35 GeV) thresholds with identification criteria allowing high signal efficiencies.

The fast commissioning of the triggers and measurements of their performance in early 2015 allowed to have first physics results promptly, many of them being also presented in these proceedings. Further modifications are on their way for 2016 aiming to bring the online algorithms even closer to the offline ones and thus further improving the performance.


\begin{thebibliography}{99}


\bibitem{ATLAS}
 ATLAS Collaboration,
  The ATLAS Experiment at the CERN Large Hadron Collider,
  \href{http://iopscience.iop.org/article/10.1088/1748-0221/3/08/S08003/}{JINST {\bf 3} (2008) S08003}.

\bibitem{ATLASElectron}
ATLAS Collaboration,
Electron efficiency measurements with the ATLAS detector using the 2012 LHC proton-proton collision data,
ATLAS-CONF-2014-032, \href{https://atlas.web.cern.ch/Atlas/GROUPS/PHYSICS/CONFNOTES/ATLAS-CONF-2014-032/}{\small \tt https://atlas.web.cern.ch/Atlas/GROUPS/PHYSICS/CONFNOTES/ATLAS-CONF-2014-032/}. 

\bibitem{ATLASPhoton}
ATLAS Collaboration,
Measurements of the photon identification efficiency with the ATLAS detector using 4.9 fb$^{-1}$ of pp collision data collected in 2011, 
ATLAS-CONF-2012-123, \href{https://atlas.web.cern.ch/Atlas/GROUPS/PHYSICS/CONFNOTES/ATLAS-CONF-2012-123/}{\small \tt https://atlas.web.cern.ch/Atlas/GROUPS/PHYSICS/CONFNOTES/ATLAS-CONF-2012-123/}.

\bibitem{ATLASTrigger}
ATLAS Collaboration,
Performance of the ATLAS Trigger System in 2010,
Eur.\ Phys.\ J.\ C {\bf 72} (2012) 1849,
\href{http://arxiv.org/abs/1110.1530}{arXiv:1110.1530 [hep-ex]}.

\bibitem{ATLASEgammaTrigger}
ATLAS Collaboration,
Performance of the Electron and Photon Trigger in p-p Collisions at $\sqrt{s}=7$ TeV with the ATLAS Detector at the LHC,
ATLAS-CONF-2011-114, \href{https://atlas.web.cern.ch/Atlas/GROUPS/PHYSICS/CONFNOTES/ATLAS-CONF-2011-114/}{\small \tt https://atlas.web.cern.ch/Atlas/GROUPS/PHYSICS/CONFNOTES/ATLAS-CONF-2011-114/}. 

\bibitem{ATLASEgammaTriggerPrelim}
ATLAS Collaboration,
Public Egamma Trigger Plots for Collision Data, 
\href{https://twiki.cern.ch/twiki/bin/view/AtlasPublic/EgammaTriggerPublicResults}{\small \tt https://twiki.cern.ch/twiki/bin/view/AtlasPublic/EgammaTriggerPublicResults}.

\end{thebibliography}
\end{document}